# HYPERPARAMETER TUNING-BASED OPTIMIZED PERFORMANCE ANALYSIS OF MACHINE LEARNING ALGORITHMS FOR NETWORK INTRUSION DETECTION

Sudhanshu Sekhar Tripathy and Bichitrananda Behera

Department of Computer Science and Engineering, C.V. Raman Global University, Bhubaneswar, Odisha, India

## ABSTRACT

*Network Intrusion Detection Systems (NIDS) are essential for securing networks by identifying and mitigating unauthorized activities indicative of cyberattacks. As cyber threats grow increasingly sophisticated, NIDS must evolve to detect both emerging threats and deviations from normal behavior. This study explores the application of machine learning (ML) methods to improve the NIDS accuracy through analyzing intricate structures in deep-featured network traffic records. Leveraging the 1999 KDD CUP intrusion dataset as a benchmark, this research evaluates and optimizes several ML algorithms, including Support Vector Machines (SVM), Naïve Bayes variants (MNB, BNB), Random Forest (RF), k-Nearest Neighbors (k-NN), Decision Trees (DT), AdaBoost, XGBoost, Logistic Regression (LR), Ridge Classifier, Passive-Aggressive (PA) Classifier, Rocchio Classifier, Artificial Neural Networks (ANN), and Perceptron (PPN). Initial evaluations without hyper-parameter optimization demonstrated suboptimal performance, highlighting the importance of tuning to enhance classification accuracy. After hyper-parameter optimization using grid and random search techniques, the SVM classifier achieved 99.12% accuracy with a 0.0091 False Alarm Rate (FAR), outperforming its default configuration (98.08% accuracy, 0.0123 FAR) and all other classifiers. This result confirms that SVM accomplishes the highest accuracy among the evaluated classifiers. We validated the effectiveness of all classifiers using a tenfold cross-validation approach, incorporating Recursive Feature Elimination (RFE) for feature selection to enhance the classifiers accuracy and efficiency. Our outcomes indicate that ML classifiers are both adaptable and reliable, contributing to enhanced accuracy in systems for detecting network intrusions.*

## 1. INTRODUCTION

The rapid growth of digital technology has improved efficiency and connectivity but also intensified sophisticated cyber threats such as ransomware, phishing, and DoS attacks. With over 90% of critical operations relying on online platforms, ensuring the confidentiality, integrity, and availability of digital assets is vital. To address these challenges, researchers are developing advanced Network Intrusion Detection Systems (NIDS) using machine and deep learning for real-time anomaly detection and proactive defense. The integration of Artificial Intelligence (AI), especially Machine Learning (ML), has greatly enhanced Network Intrusion Detection Systems (NIDS). These systems analyze network traffic to distinguish normal and malicious activities, detecting zero-day attacks that evade traditional defenses. According to NIST, intrusion detection ensures data confidentiality, integrity, and availability through continuous monitoring and anomaly detection [1]. Unlike signature-based systems, anomaly-based NIDS effectively identify previously unseen attacks by learning normal traffic behavior.

Recent advances in machine learning (ML) and deep learning (DL) have enhanced IDS performance through adaptive, data-driven detection with fewer false positives [2] [3]. Techniques like SVM, RF, DT, KNN, XGBoost, AdaBoost, and BPN are widely used, while CNNs and RNNs capture complex traffic patterns [4]. Hyperparameter tuning and feature engineering optimize accuracy and scalability [5] [6]. Using the KDD Cup 1999 dataset, this study evaluates multiple classifiers based on accuracy, precision, recall, F1-score, false alarm rate, and detection rate to strengthen IDS robustness.

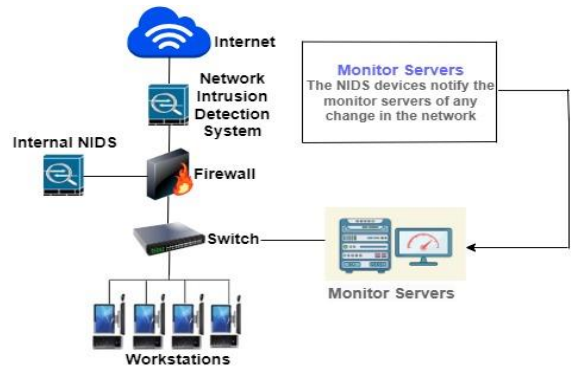

Fig.1. A snapshot of systems for detecting network intrusions

Fig. 1 illustrates the system layout of a system for detecting network intrusions, engineered to observe and assess network traffic, identifying potential intrusions or suspicious activities. The system connects to the internet through a firewall that filters traffic according to established security rules. NIDS sensors are positioned at both external and internal points to examine network packets. Traffic is routed through a switch that connects various workstations within the network. When anomalies or suspicious behavior are detected, the NIDS sends alerts to monitoring servers, which evaluate the threat's severity and manage response actions. This centralized setup enables continuous surveillance, strengthening the network's defense against cyber-attacks by allowing real-time threat detection and proactive response.

This research highlights the following major contributions:

- The initial investigation establishes baseline performance metrics for machine learning classifiers applied to systems for detecting network intrusions using the 1999 KDD Cup intrusion detection dataset without hyper-parameter tuning. A diverse range of ML algorithms was systematically assessed through tenfold cross-validation without applying any feature selection techniques. This analysis provides critical perspectives on their performance, highlighting capabilities and restrictions and emphasizing the need for further optimization to improve accuracy and lower false alarm rates. The findings serve as a valuable benchmark for guiding future advancements in network security solutions.
- The second investigation applies sophisticated hyper-parameter optimization strategies like grid search and random search to boost the performance of ML classifiers. Subsequently, all classifiers were evaluated through tenfold cross-validation with RFE, improving accuracy, efficiency, and emphasizing essential features. The results clearly show that systematic tuning of hyper-parameter configurations leads to significant improvements, enhancing detection accuracy while minimizing the rate of incorrect positive detections.

This investigation emphasizes the importance of hyper-parameter optimization in improving the durability and trustworthiness of systems for detecting network intrusions, contributing to facilitating the progress of more effective and efficient classifiers for practical deployment in network security applications.

Section 2 presents a comprehensive investigation into recent advancements in systems for detecting network intrusions, emphasizing a critical analysis of methodologies, emerging trends, prevailing challenges, and a systematic comparative evaluation of relevant research studies. Section 3 outlines an enhanced framework for detecting network intrusions utilizing the 1999 KDD intrusion detection dataset. It provides a detailed depiction of a machine learning-based NIDS architecture, emphasizing its procedural framework components and optimization strategies. Section 4 presents an ML-driven design for detecting network intrusions, emphasizing the impact of prominent ML classifiers in improving detection efficiency and ensuring comprehensive performance evaluation. Section 6 clarifies the experimental setup and performance evaluation executed utilizing the 1999 KDD Cup intrusion dataset. It includes an analysis of confusion matrices, hyper-parameter tuning, and a comparative analysis of results before and after optimization, emphasizing the influence of hyper-parameter optimization on accuracy and false positive rates. Section 7 wraps up by outlining the significant outcomes of the study and proposing future avenues of research to advance next-generation NIDS.

## 2. RELATED WORK

A study [7] employed an XGBoost-based feature selection approach, identifying 17 and 22 optimal features for the NSL-KDD and UNSW-NB15 datasets, respectively. The XGBoost-LSTM hybrid achieved 99.49% validation accuracy and 88.13% test accuracy on NSL-KDD, while XGBoost-Simple-RNN attained 87.07% on UNSW-NB15. Another study [8] introduced HCRNNIDS, a hybrid CRNN integrated with logistic regression, decision trees, and XGBoost, achieving 97.75% accuracy on CSE-CIC-DS2018 and outperforming several traditional and deep learning IDS models. In [9], a hybrid anomaly detection model integrating a classical autoencoder (CAE) with a deep neural network (DNN) was applied to the UNSW-NB15 dataset. The CAE enhanced DNN performance through sparse feature extraction, achieving 91.29% accuracy and outperforming baseline models. Similarly, [10] proposed a deep learning-based IDPS using an MLP trained on KDD CUP 1999, optimized with Adam, achieving 91.4% accuracy compared to DT (74%) and SVM (83%) classifiers.

In [11], a hybrid IoT intrusion detection model combined random forest-based feature selection with neural classifiers (B-ANN and DR-NN), achieving 98% accuracy on KDD CUP 1999 and demonstrating strong adaptability across intelligent networks. Similarly, [12] evaluated NB, DT, KNN, RF, SVM, MLP, and LSTM on NSL-KDD, reporting accuracies of 89.6% (with scaling), 89.2% (without), 96.89% (MLP), and 97.77% (LSTM), confirming LSTM's superiority in modeling temporal dependencies. The study in [13] highlighted CNNs as highly effective for IoT intrusion detection, demonstrating deep learning's advantage over traditional methods. In [14], DNN and LSTM models on NSL-KDD showed that a three-layer LSTM with 32 neurons per layer achieved 98.3% accuracy, outperforming enlarged DNNs and conventional models. Reference [15] applied a deep autoencoder for five-class IDS on NSL-KDD, achieving 99% training and 91.28% testing accuracy. In [16], an LSTM-based IDS on CIDDS reached 0.85 accuracy, surpassing SVM, MLP, and Naïve Bayes.

A CNN-IDS in [17] applied dimensionality reduction on KDD 1999 data, converting traffic into image-like representations to reduce complexity. Results demonstrated higher accuracy and lower FAR compared with conventional methods. The study in [18] proposed a deep belief network (DBN) framework optimized with PSO, clustering, and genetic operators, reducing detection time by 24.69% and improving five-class accuracy by 1.3–14.8%. An improved LeNet-5 CNN in [19] integrated normalization and one-hot encoding, achieving over 99% training and evaluation accuracy with FAR below 0.1%, emphasizing reliability and precision. DL-IDS in [20] combined CNN and LSTM for feature extraction, with category weight optimization to handle class imbalance. On CICIDS2017, multi-class accuracy reached 98.67%, with over 99.5% for certain attack classes, showing its effectiveness for diverse intrusion patterns. In [21], a DBN-ELM hybrid applied feature extraction and classification on NSL-KDD, using majority voting to refine predictions, achieving 97.82% accuracy and a 1.81% false alarm rate, outperforming individual DBN or ELM models. The deep multilayer framework in [22], incorporating feedback, autoencoding, preprocessing, database management, and classification, attained 96.70% accuracy on NSL-KDD, highlighting the advantage of integrated architectures. In [23], a stacked nonsymmetric deep autoencoder (NDAE) enhanced unsupervised feature extraction on KDD 1999 and NSL-KDD, significantly improving detection performance over traditional NIDS.

Table 1. Comparative Analysis of Related Works on Network Intrusion Detection Systems

| Reference | Data set | Classifiers Applied | Detected Assaults | Evaluated Matrix With Accuracy | Findings |
|---|---|---|---|---|---|
| Kasongo, S. M [7] | NSL KDD, UNSW-NB15 | XGBoost-LSTM, XGBoost-Simple-RNN, XGBoost-GRU | Dos, Probe, R2L, U2R, Normal Normal, Generic, Exploits, Fuzzers, DoS, Reconnaissance, Analysis, Backdoor, Shellcode, Worms | F1-Score, TAC, VAC Accuracy XGBoost-LSTM (TAC) = 88.13% and VAC of 99.49% (NSL KDD) For Multiclass Classification, XGBoost –LSTM (TAC) = 86.93% (NSL KDD) XGBoost-Simple-RNN (TAC) = 87.07% (UNSW NB15) XGBoost-GRU (TAC) = 78.40% (UNSW NB15) | XGBoost-LSTM Model Performance • Outperformed other approaches with TAC of 88.13%. • An assessment proficiency of 99.49%. • A training period of 225.46 seconds for binary classification tasks. |
| Khan, M.A. [8] | CSE-CIC-DS2018 | LR, XGB, DT, HCRNN | Brute-force DOS attacks, DDOS attacks, Brute-force SSH, Infiltration, Heartbleed, Web attacks, and Botnet. | FP, TP, FN, TN Prec, Rec, F1-Score, DR, FPR Accuracy LR = 80% XGB = 83% DT = 88% HCRNN = 97.75% | HCRNNIDS Deep Learning Model Simulation Results • Accurately calculates malicious attack events. • Overall accuracy: 97.75%. • Effective security solution. |
| Dutta, Vibe kananda & Pawlicki, Mare | UNSW-NB15 | RF,DNN,Hybrid (CAE+DNN) | Normal, Generic, Exploits, Fuzzers, DoS, Reconnaissance, Analysis, | Prec, Rec, Acc, F1-score, FPR, ROC curve Accuracy RF = 85.14% DNN = 88.15% Hybrid (CAE+DNN) = 91.29% | Hybrid Approach Performance • Superior in distinguishing attacks from routine activities. • Comparable to other baseline algorithms. |

| | | Backdoor, Shellcode, Worms | | |
|---|---|---|---|---|
| k et al. [9] | | | | |
| Akhil Krishna, Ashik Lal M.A, et al. [10] | KDD CUP' 99 | DT, SVM, MLP | DOS, Probe, U2R, R2L and Normal | Accuracy DT = 74.63% SVM = 83.06% MLP = 91.41% | Deep Learning MLP Model Improvement • Improved accuracy to 91.41%. • Completed intrusion detection system model. • Utilized sparse categorical cross-entropy loss function. |
| M.Ramaiah, V. Chandrasekaran et al. [11] | KDD CUP' 99 | Proposed S-NN, D-ONN | Normal, DOS, Probe, U2R, R2L | TP, TN, FP, FN Acc, Prec, Rec, F1-Score Accuracy S-NN (Shallow neural network model) =96% D-ONN (Deep-optimized neural network) = 98% | Intrusion Detection Framework: 98% Accuracy" • Utilizes correlation tools and Random Forest. • Focuses on cyber-physical system IDS. |
| Hossain, Md & Ghose, Dipayan et al. [12] | NSL KDD | MLP,LSTM,NB,DT,KNN,RF, SVM | Normal, Dos, Probe, ,R2L, U2R | Acc, Prec, Rec, and F1-Score Accuracy LSTM = 97.77% MLP = 96.89% NB = 75.9% DT = 88.2% KNN = 87.0% RF = 89.6% SVM = 87.6% | NSL KDD dataset accuracy • LSTM (97.77%) and MLP (96.89%) has been implemented. • The dataset consists of two labels and 41 traffic-related input features for each record. |
| Jose, Jinsi & Jose, Deepa [13] | NSL KDD | DT, DNN, CNN, | Dos, Probe, ,R2L, U2R | Acc, Prec, Rec and F1-Score, FPR, TPR Accuracy DT = 80% DNN = 86% CNN = 89% | DL for IDS • Networks of convolutional neurons show 89% accuracy. • High-prediction assault detection. |
| Zarai, R., Kachout, M. et al. [14] | KDD CUP' 99 | Proposed LSTM and DNN | Normal, Dos, Probe, ,R2L, U2R | Acc, Prec, Rec, and F1-Score Accuracy LSTM = 98.3% DNN = 93% | Three-Layer LSTM Outperforms Traditional Machine Learning • Accuracy: 98.3% |
| B. Alsughayyir, A. M. Qamar et al. [15] | NSL KDD | Deep Auto Encoder | Dos, Probe, ,R2L, U2R,Normal | Prec, Rec, F1-Score, Support Accuracy Deep Auto Encoder = 99.90% | Proposed DL Strategy Outperforms Traditional Methods • 99% training accuracy • 91.28% testing accuracy. |
| S. A. Althubiti, E. M. Jones, et al. [16] | CID DS-001 | LSTM, SVM, NB, MLP | probes, user-to-root, remote-to-local attacks | Acc, Prec, Rec, FPR Accuracy LSTM = 0.8483 SVM = 0.7942 NB = 0.7756 MLP = 0.8124 | LSTM Model Outperforms SVM, MLP, Naïve Bayes in Multiclassification • Achieves acceptable accuracy of 0.8483. |
| Y. Xiao, C. Xing, T.et al. [17] | KDD CUP' 99 | CNN-IDS | Dos, Probe, ,R2L, U2R,Normal | ACC, DR, FAR Accuracy CNN-IDS = 94% | CNN-IDS Model: 94% Timeliness, FAR, AC Outperformance • Beneficial for research and real-world applications. |
| P. Wei, Y. Li, Z. Zhang et al. [18] | NSL KDD | DBN (Deep Belief Network) | Dos, Probe, ,R2L, U2R | Accuracy, FNR, FPR, DR Accuracy DBN = 82.36% | DBN-IDS Model Optimization • Achieved 82.36% detection accuracy. • Demonstrated faster detection speed. |
| Liu, Pengju [19] | KDD CUP' 99 | CNN | Dos, Probe, ,R2L, U2R,Normal | ACC, DR, FPR Accuracy CNN = 99% | Model's High Testing, Training, and Detection Accuracy • DR less than 0.1% • Performs well in actual detection tests. |
| Sun, Pengfei & Liu, Pengju eta al. [20] | CICI DS2017 | hybrid model using CNN and LSTM | Brute Force FTP, Brute Force SSH, DoS, Heartbleed, Web Attack, Infiltration, Botnet, and DDoS | ACC, TPR, FPR, Prec, Rec, F1-Score Accuracy CNN-LSTM = 98.67% | DL-IDS Outperforms Machine Learning Models • Achieves 98.67% accuracy • Achieves 93.32% F1-score. |
| D. Liang and P. Pan [21] | NSL KDD | DBN-ELM | Dos, Probe, ,R2L, U2R | Accuracy DBN-ELM = 97.82% | Model Accuracy Enhancement: 97.82% • Reduced false alarm rates to 1.81%. • Attained competitive accuracy compared to DBN, ELM, DBN-ELM. |
| Ugen | NSL | Deep | Dos, | Accuracy | Deep Multilayer |

| dhar, A. & Illuri, Babu at al. [22] | KDD | multilay er classific ation network | Probe, ,R2L, U2R | Deep multilayer classification = 96.70% | Classifier Performance • Outperformed all methods in accuracy. • Achieved 96.70% score in comparative results. |
|---|---|---|---|---|---|
| N. Shon e, T. N. Ngoc, V. D. Phai, et al. [23] | KDD CUP' 99 and NSL KDD | Non-symmet ric deep auto-encoder (NDAE ) | Normal, Dos, Probe, ,R2L, U2R | Accuracy KDD CUP'99 = 97.85% NSL KDD = 85.42% | Evaluation of the framework by NDAE • Offers high acc, prec, and rec. • Reduces training time. • Comparatively compared with mainstream DBN technique. • Shows up to 5% accuracy improvement and 98.81% training time reduction. • Model's capabilities across both reference datasets. • Steady classification performance. |

## 3. AN OPTIMAL APPROACH TO SYSTEMS FOR DETECTING NETWORK INTRUSION APPLYING THE 1999 KDD CUP DATASET

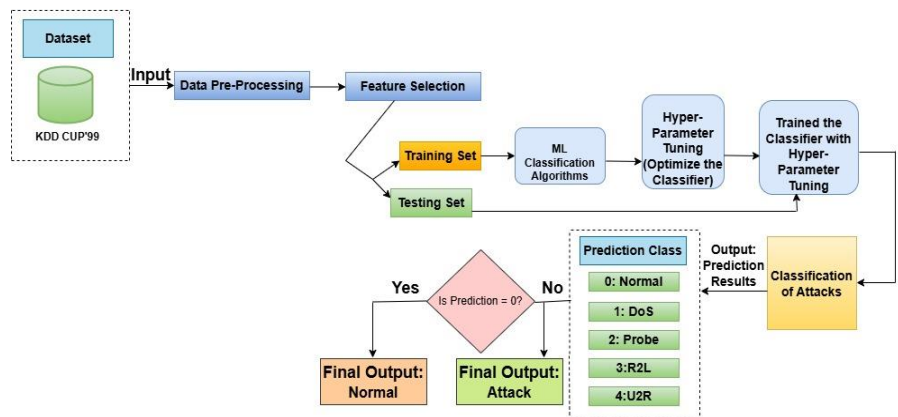

Fig. 2 ML Based NIDS architecture

Fig. 2 depicts the optimal workflow for NIDS using the 1999 KDD Cup intrusion dataset, which involves several well-defined steps to support accurate and timely detection and classification of cyber intrusions targeting network infrastructure:

➢ **Dataset Utilization**: The 1999 KDD Cup intrusion dataset serves as the foundational input applicable to the NIDS architecture. It is broadly adopted for detecting malicious intrusions in networks, offering a diverse range of assigning network traffic data to normal activity or specific cyberattack classifications.

➢ **Data Pre-Processing**: The raw data undergoes pre-processing to enhance its quality. This step includes handling missing values, eliminating redundant logs, and applying normalization or scaling to features. Pre-processing verifies that the dataset remains accurate and prepared for evaluation.

➢ **Feature Selection**: Before model training, Recursive Feature Elimination (RFE) was utilized to optimize the feature space, ensuring that only the most relevant attributes were used in classification. By iteratively eliminating less significant features, RFE streamlines the dataset, which leads to faster processing times and improved model accuracy. This targeted approach allows classifiers to focus on essential indicators of network intrusions, thereby strengthening their detection capabilities.

➢ **Data Splitting**: The dataset is divided into separate segments for training and evaluating the classifier. The training set is used to build and fine-tune the ML classifiers, while the testing set assesses how well the trained model performs and generalizes to unseen data.

➢ **Classifier Training with ML Algorithms**: Multiple ML algorithms are trained on the dataset to uncover patterns and correlations in the network traffic. This allows the models to accurately classify traffic as either normal or belonging to specific types of attacks.

➢ **Hyper-parameter Optimization**: Optimization of hyper-parameters is implemented to fine-tune the classifier's accomplishment. This process includes fine-tuning parameters like the learning rate, the size of estimators in ensemble methods, or the depth of decision trees to achieve the best possible results.

➢ **Trained Classifier:** Using the optimal hyper-parameters, the classifier is modeled using the training data inputs. This yields a classifier capable of accurately forecasting the category of novel, unobserved occurrences based on learned patterns.

➢ **Multi-Class Prediction**: The trained classifier generates predictions for each instance, assigning them to one of the following categories:

> 0: Normal Activity
> 1: Denial-of-Service (DoS) Attack
> 2: Probing/Scanning Attempt
> 3: Remote-to-Local (R2L) Intrusion
> 4: User-to-Root (U2R) Privilege Escalation

➢ **Decision Block (Normal or Attack):** A decision block is implemented to verify whether the prediction corresponds to the "Normal" class (prediction = 0). If the prediction equals 0, the instance is classified as normal. Otherwise, if the prediction matches any attack class, the instance is categorized as an attack

➢ **Attack Classification**: For instances categorized as attacks, the system further classifies them into specific attack types such as DoS, probe, R2L, or U2R. This fine-grained classification enables precise identification and differentiation of attack types within the broader category of malicious activities.

## 4. NETWORK INTRUSION DETECTION SYSTEMS WITH ML CLASSIFIERS

### 4.1. Classifiers and Techniques in Machine Learning

Machine learning enhances NIDS by enabling autonomous intrusion detection through data-driven pattern recognition [24], [25]. Supervised learning offers high accuracy using labeled data [26], while unsupervised learning detects anomalies without labels but with lower accuracy [27]. Both approaches improve NIDS performance, strengthen security, and reduce false positives [28].

## 4.2. Classification Approach Using Support Vector Machines

SVMs are widely used in NIDS for their high accuracy in detecting and classifying network anomalies. They classify data by finding a maximum-margin separator between normal and malicious traffic, relying on support vectors for efficiency even with limited training data. For non-linear patterns, kernel methods map inputs to higher-dimensional spaces, enabling complex decision boundaries [29]. This approach minimizes classification errors and false positives, making SVMs robust and versatile for both linear and non-linear intrusion detection scenarios.

## 4.3. Probabilistic Learning Classifier Using Naïve Bayes

The Naive Bayes classifier, based on Bayes' theorem, predicts class probabilities by assuming conditional independence among features. Variants such as Multinomial Naive Bayes (MNB) handle count data, while Bernoulli Naive Bayes (BNB) processes binary features. In NIDS, it is valued for simplicity, scalability, and computational efficiency, enabling effective analysis of high-dimensional network data. Despite the strong independence assumption, Naive Bayes reliably differentiates normal from malicious connections, providing a lightweight intrusion detection solution [30].

$$P(S|T) = \frac{P(T|S) * P(S)}{P(T)} \tag{1}$$

Where:

    T: Observed features or data.
    S: Target class or category.
    P(S|T): Probability of class S given data T.
    P(S): Prior probability of class S.
    P(T|S): Probability of data T given class S.
    P(T): Overall probability of data T.

## 4.4. Classification Technique Using a Decision Tree

Decision trees (DTs) are a popular supervised learning method for classification and regression, structured as hierarchical trees with internal nodes for feature-based decisions, branches for outcomes, and leaves for predictions. In NIDS, DTs effectively detect normal and malicious traffic using features such as connection duration, protocol, and service type. Their interpretability and feature-driven decision process allow efficient handling of complex datasets, providing accurate and real-time intrusion detection with computational efficiency in large-scale networks [31].

## 4.5. K-Nearest Neighbor based Classification Technique

K-Nearest Neighbors (KNN) is a non-parametric, distance-based, instance-based learning method widely used in NIDS for its simplicity and effectiveness. It classifies a data point based on the majority label among its K nearest neighbors, using metrics such as Euclidean distance. By comparing network connections with labeled training instances, KNN identifies normal and malicious patterns. Although computationally intensive for large datasets, techniques like dimensionality reduction and approximate neighbor search enhance its scalability and efficiency [32].

## 4.6. Classification Approach Using Logistic Regression

Logistic Regression (LR) is a supervised algorithm used in intrusion detection to classify network traffic as normal or malicious. It applies the logistic function to generate outputs between 0 and 1, estimating the probability of each class and making predictions based on a threshold. LR is efficient, interpretable, and computationally lightweight, providing probabilistic predictions. However, its simplicity may limit performance on complex, high-dimensional data, where more advanced models often perform better [33].

In logistic regression, a linear model is derived from the provided attributes and processed through a sigmoid curve, resulting in a probabilistic output. The sigmoid function is mathematically expressed as:

$$F(x) = 1 / 1 + e\text{-}x \qquad\qquad (2)$$

In this equation, F(x) yields a probability between 0 and 1, with "e" standing for the natural exponential base, and "x" acting as the function's input.

## 4.7. Classifier Using Linear Discriminant Analysis Technique

Linear Discriminant Analysis (LDA) is a supervised method used in intrusion detection to classify network traffic and reduce feature dimensionality. It maximizes differences between classes while minimizing within-class variance, identifying linear combinations of features that enhance separability. LDA effectively classifies traffic into normal or specific attack types, supports multi-class detection, and improves computational efficiency by preserving class separability in lower-dimensional space [34].

## 4.8. Optimized Extreme Gradient Boosting (XGBOOST) Classifier

XGBoost is a scalable gradient boosting algorithm widely used in network-level intrusion detection for its efficiency with large and complex datasets. It combines multiple weak learners, typically decision trees, to iteratively improve predictive performance by correcting previous errors. This approach effectively handles high-dimensional and imbalanced data, enabling accurate detection of diverse and novel intrusion types, making XGBoost a robust solution for precise NIDS implementation [35].

## 4.9. AdaBoost Classifier

AdaBoost is a boosting algorithm commonly used in network intrusion detection for its ability to improve accuracy by combining weak learners into a strong classifier. It assigns higher weights to misclassified instances, ensuring subsequent models focus on difficult or ambiguous patterns. This adaptive approach reduces false positives and effectively handles high-dimensional, imbalanced network data, enhancing detection of normal and malicious activities, including emerging or unknown threats

## 4.10.  Random Forest Classifier

Random Forest (RF) is an ensemble learning method widely used in network-layer intrusion detection for its accuracy and robustness against overfitting. It constructs multiple decision trees on varied data subsets and aggregates their predictions, capturing complex, non-linear patterns in high-dimensional NIDS datasets. RF effectively detects both known and zero-day threats, handles imbalanced data, and ranks critical features to enhance accuracy while reducing computational demands [36].

### 4.11. Artificial Neural Network (ANN)

Artificial Neural Networks (ANNs) are widely used in network intrusion detection for their ability to model complex, non-linear data. They comprise an input layer for network features, hidden layers for feature extraction, and an output layer for classification. Neurons are interconnected with weighted links, and activation functions such as ReLU, Sigmoid, Tanh, and Softmax process inputs. Methods like Perceptron, SGD, and backpropagation optimize the network by minimizing errors. Deep ANN architectures improve detection accuracy, enhance system performance, and reduce false alarms [37].

### 4.12. Ridge Classifier

The Ridge classifier assumes that data points of each class lie within a linear subspace, enabling continuous analysis for classification [38]. In NIDS, it addresses multicollinearity among network features through L2 regularization, stabilizing predictions and reducing variance. By controlling model complexity, Ridge regression minimizes overfitting and ensures accurate, reliable detection of network anomalies, making it suitable for high-dimensional intrusion detection tasks.

### 4.13. Passive Aggressive (PA) Classifier

Passive-Aggressive (PA) classifiers are scalable online learning algorithms that update models incrementally as new data arrives, unlike traditional batch methods. In NIDS, they adapt to evolving network conditions by processing streaming data efficiently. Using a regularization parameter (C) instead of a learning rate, PA classifiers penalize misclassifications to balance accuracy and model simplicity. This enables real-time anomaly detection with low computational overhead, making them well-suited for high-traffic networks [39].

### 4.14. Rocchio (RC) Classifier

The Rocchio algorithm, originating from relevance feedback in information retrieval, is applied in NIDS for classification. During training, it computes a centroid for each class as a prototype. In testing, class labels are assigned based on the Euclidean distance between incoming data points and centroids. This proximity-based method efficiently detects anomalies, distinguishing normal traffic from potential intrusions while helping minimize false positives.

## 5. RESULTS AND DISCUSSION

### 5.1. Experimental Setup

Machine learning computations were performed using Python's Scikit-learn library. Experiments were conducted on Google Colaboratory, a cloud-based platform equipped with a Tesla K20 GPU (2,496 CUDA cores, 16 GB RAM, and 500 GB storage), as well as locally on a Windows 11 system powered by an Intel Core i5-1240P processor (4.40 GHz, 12th generation), identified as DESKTOP-UFN62J4. This dual setup facilitated a comprehensive evaluation of machine learning classifiers in both cloud and local computing environments.

## 5.2. Dataset Description

NIDS monitor network activity to identify abnormal patterns indicative of security threats while allowing normal traffic. Machine learning classifiers, trained on datasets containing both normal and attack patterns, improve detection by recognizing diverse network behaviors. In this study, the 1999 KDD Cup dataset was used, with 70% of data for training and 30% for testing to preserve class distribution. A 10-fold cross-validation was also applied to rigorously evaluate model accuracy, generalization, and robustness against overfitting.

## 5.3. The 1999 KDD CUP Intrusion Dataset

The 1999 KDD Cup intrusion dataset is a widely used benchmark for evaluating network intrusion detection mechanisms. Developed for the KDD 1999 data mining challenge, it contains simulated network traffic with both normal and malicious connections. The dataset includes approximately 4.9 million records, each described by 41 features capturing key aspects of network behavior, such as connection duration, protocol type, and error rate, making it a foundational resource for anomaly detection and machine learning in cybersecurity. The 1999 KDD Cup dataset categorizes malicious activity into four types: Denial of Service (DoS), Remote-to-Local (R2L), User-to-Root (U2R), and Probe attacks. Normal network traffic is also included to provide a baseline for training and evaluation. The dataset is typically split into training and testing subsets, with cross-validation used to assess detection performance. Despite criticisms such as data redundancy, it remains a widely accepted benchmark for developing and evaluating NIDS.

(i) **Denial-of-Service (DoS) Assaults**: These types of exploits focus on disrupting network infrastructure or system operations by inundating them with excessive traffic or requests, such as in a SYN Flood attack.

(ii) **Remote-to-Local (R2L) Intrusion**: These occur when a malicious agent gains access to a local computer remotely without valid credentials, often through techniques like password cracking.

(iii) **Probing/Scanning Attempt**: These involve reconnaissance activities aimed at collecting information about a network's structure and identifying vulnerabilities, such as through port scanning.

(iv) **User-to-Root (U2R) Privilege Escalation**: In these breach attempts, an intruder attempts to elevate privileges from a regular user account to administrator (root) access, often using methods like buffer overflow exploits.

Fig 3. Feature and Label Structure of the 1999 KDD Cup intrusion detection dataset

Fig. 3 illustrates the process of loading and displaying the 1999 KDD Cup dataset using Python's pandas library. The CSV files, kddtrain.csv and kddtest.csv, are imported into DataFrames traindata and testdata, with header=None indicating no header row. The command traindata. head (8) displays the first eight rows, showing 42 columns indexed from 0 to 41. Each row represents a network connection, and each column corresponds to attributes such as protocol type, connection duration, and status.

The 1999 KDD Cup dataset is widely used for detecting malicious network activity and includes four types of features: basic, content, traffic, and class labels, as summarized in Table 2. Basic features describe connection properties, such as duration, protocol type (TCP, UDP, ICMP), service (HTTP, FTP), flags, and data transfer metrics (src_bytes, dst_bytes). Content features capture connection-level activities, including failed logins, user login status, and system-level actions like root_shell or su_attempted. Traffic features aggregate session details, such as number of shells, accessed files, and login types (host_login, guest_login). Class labels distinguish normal traffic from attacks, including DoS, R2L, U2R, and Probe. These features are critical for training machine learning classifiers for effective intrusion detection.

Table 2. Highlights the frequency distribution of cases among multiple attack classes in the 1999   KDD Cup intrusion dataset

| Sets | Traffic Categories | Authentic Logs | Unique Data Points |
|---|---|---|---|
| Training Set | Intrusions | 3,925,650 | 262,178 |
| | Benign Traffic | 972,781 | 812,814 |
| | Overall Count | 4,898,431 | 10,74,992 |
| Testing Set | Intrusions | 2,46,150 | 29,378 |
| | Benign Traffic | 60,591 | 47,911 |
| | Total | 306,741 | 77,289 |

## 5.4. Performance Evaluation Metrics of NIDS

Evaluating network monitoring and intrusion detection systems is crucial for enhancing threat detection, refining algorithms, reducing false positives, and ensuring operational reliability. Performance is measured using metrics such as Accuracy (Acc), Precision (Prec), Recall (Rec), F1-score, False Alarm Rate (FAR), and Detection Rate (DR). Table 3 summarizes four key outcomes: true positives (TP), false positives (FP), true negatives (TN), and false negatives (FN), which form the basis for performance assessment. A confusion matrix organizes these outcomes, allowing computation of the key metrics and providing a structured framework for evaluating machine learning classifiers in intrusion detection.

**True Positive (TP)**: An intrusion attempt is correctly recognized by the system as malicious, confirming successful threat detection.

**False Positive (FP)**: Benign traffic is incorrectly flagged as a threat, triggering an unnecessary alert.

**True Negative (TN)**: Safe network activity is accurately classified as non-malicious, resulting in no false warning.

**False Negative (FN)**: A harmful activity passes through undetected and is wrongly classified as legitimate, signifying a lapse in the detection mechanism.

Table 3. Calculating NIDS Performance Metrics

| Actual \ Predicted | Attack (Positive) | Normal (Negative) |
|---|---|---|
| **Attack (Positive)** | True Positive (TP) | False Negative (FN) |
| **Normal (Negative)** | False Positive (FP) | True Negative (TN) |

**Accuracy**: Accuracy measures the proportion of correctly classified normal and attack instances, offering an overall evaluation of NIDS performance across all predictions. Mathematically:

$$\text{Accuracy}^{\text{NIDS}} = \frac{TP + TN}{TP + FP + TN + FN} \tag{3}$$

**Precision**: Precision indicates NIDS reliability in detecting attacks. It reflects how accurately alerts are raised, with higher precision values representing fewer false alarms:

$$\text{Precision}^{\text{NIDS}} = \frac{TP}{TP + FP} \tag{4}$$

**Recall**: Recall measures how effectively NIDS detects actual attacks. It reflects the system's ability to capture malicious activities without missing threats, ensuring comprehensive detection coverage:

$$\text{Recall}^{\text{NIDS}} = \frac{TP}{TP + FN} \tag{5}$$

**F1 Score**: F1-score balances precision and recall through their harmonic mean. It is especially useful for NIDS evaluation on imbalanced datasets, ensuring neither metric dominates performance assessment.

$$\text{F1 Score}^{\text{NIDS}} = \frac{2(recall * precision)}{recall + precision} \tag{6}$$

**False Alarm Rate:** False Alarm Rate in NIDS measures the frequency of normal traffic misclassified as attacks. Lower values indicate higher reliability and reduced unnecessary security alerts.

$$\text{FAR}^{\text{NIDS}} = \frac{FP}{FP + TN} \tag{7}$$

## 5.5. Confusion Matrices for ML Classifiers on the 1999 KDD Cup Intrusion Dataset

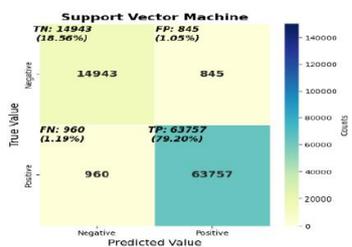 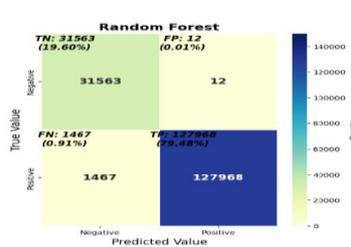 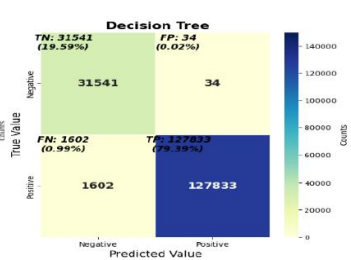

| Fig 4. Confusion Matrix representing SVM | Fig 5. Confusion Matrix representing RF | RFFig 6. Confusion Matrix representing DT |
|---|---|---|

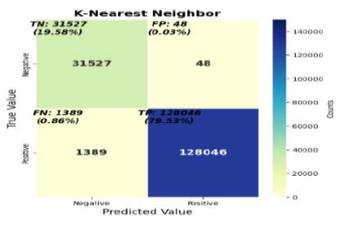

Fig 7. Confusion Matrix representing KNN

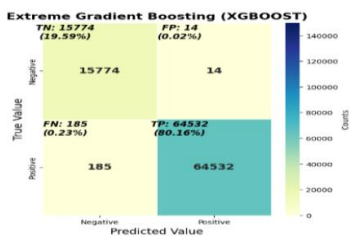

Fig 8. Confusion Matrix representing XGBOOST

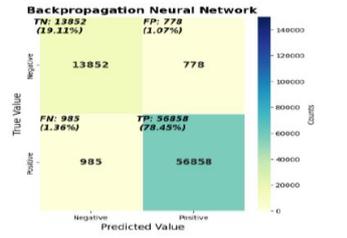

Fig 9. Confusion Matrix representing BPN

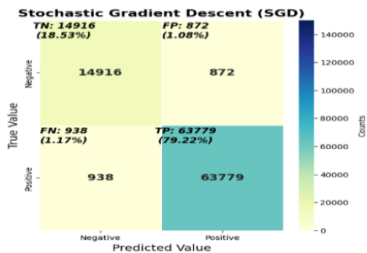

Fig 10. Confusion Matrix representing SGD

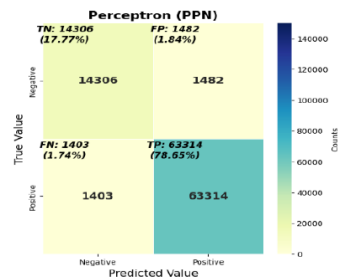

Fig 11. Confusion Matrix representing PPN

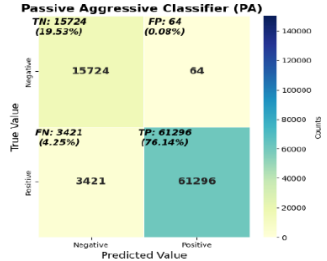

Fig 12. Confusion Matrix representing PA

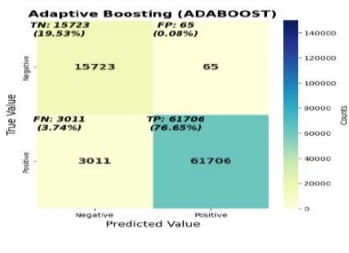

Fig 13. Confusion Matrix representing ADABOOST

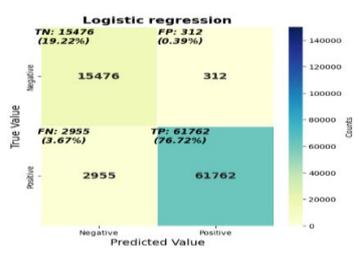

Fig 14. Confusion Matrix representing LR

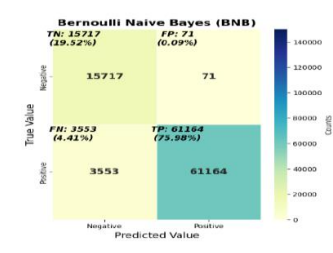

Fig 15. Confusion Matrix representing BNB

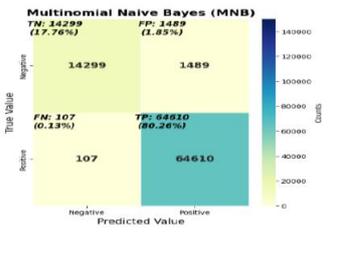

Fig 16. Confusion Matrix representing MNB

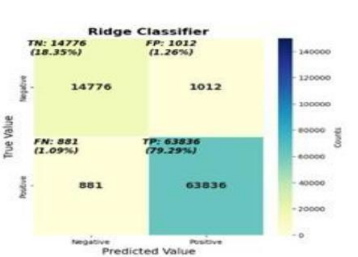

Fig 17. Confusion Matrix representing Ridge

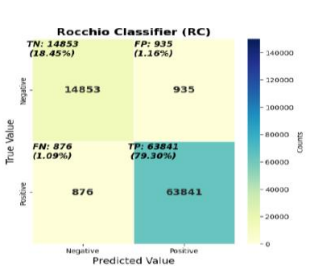

Fig 18. Confusion Matrix representing RC

## 5.6. Hyper-Parameter Tuning

In the experimental setup shown in Table 4, hyper-parameter optimization is employed to improve NIDS detection accuracy by fine-tuning key parameters. Techniques such as Grid Search and Random Search systematically explore parameter combinations to identify optimal settings. Cross-validation assesses classifier generalization across datasets, mitigating overfitting and enhancing reliability in detecting malicious activities. This tuning ensures effective real-world performance, balancing accurate detection with minimal false alarms.

Table 4. Setup of Hyper-parameters for Different Classification Techniques

| Classifiers | Parameters | | | |
|---|---|---|---|---|
| KNN | **K=10 (Best value)** K= (1,2,3,4,5,6,7,8,9,10,11,12,13,14,15) | Metric= "minkowski" | weights= "distance" | algorithm= "auto" |
| DT | **Splitting= "Gini" (Best value)** Splitting= "Entropy" Splitting= "Gini" Splitting= "Entropy" Splitting= "Gini" | splitter= "best" splitter= "best" splitter= "random" splitter= "best" splitter= "random" | min_samples_split=2 min_samples_split=5 min_samples_split=4 min_samples_split=10 min_samples_split=8 | random state=3 random state=42 random state=7 random state=0 random state=12 |
| MNB | **alpha=0.01 (Best value)** alpha=0.1 alpha=0.5 alpha=1.0 alpha=0.001 | fit_prior=True fit_prior=True fit_prior=True fit_prior=True fit_prior=True | class_prior=None class_prior=None class_prior=None class_prior=None class_prior=None | force_alpha=True force_alpha=True force_alpha=True force_alpha=True force_alpha=True |
| BNB | **alpha=0.01 (Best value)** alpha=0.05 alpha=0.001 alpha=0.1 alpha=0.5 | binarize= "1.0" binarize= "1.0" binarize= "1.0" binarize= "1.0" binarize= "1.0" | fit_prior=True fit_prior=True fit_prior=True fit_prior=True fit_prior=True | class_prior= None class_prior= None class_prior= None class_prior= None class_prior= None |
| RF | **n_estimators= "200" (Best value)** n_estimators= "100" n_estimators= "150" n_estimators= "300" n_estimators= "250" | Splitting= "gini" Splitting= "entropy" Splitting= "gini" Splitting= "gini" Splitting= "entropy" | min_samples_split=2 min_samples_split=2 min_samples_split=3 min_samples_split=4 min_samples_split=2 | max_features= "None" max_features= "sqrt" max_features= "auto" max_features= "None" max_features= "sqrt" |
| SVM | **C= 1.0 (Best value)** C= 0.5 C= 1.5 C= 0.8 | Penality factor= "l2" Penality factor= "l1" Penality factor= "l2" Penality factor= "l1" | tolerance(tol)="1e-4" tolerance(tol)="1e-3" tolerance(tol)="1e-4" tolerance(tol)="1e-3" | Loss= "square_hinge" Kernel= "rbf" Loss= "square_hinge" Kernel= "rbf" Loss= "hinge" Kernel= "linear" Loss= "square_hinge" Kernel= "sigmoid" |
| PPN | **max_iter= "150" (Best value)** max_iter= "100" max_iter= "200" | Penalty= "elasticnet" Penalty= "l2" Penalty= "elasticnet" | tolerance(tol)= "1e-3" tolerance(tol)= "1e-4" tolerance(tol)= "1e-3" | n_iter_no_change= "20" n_iter_no_change= "10" n_iter_no_change= "15" |
| **LR** | **max_iter= "150" (Best value)** max_iter= "200" | solver= "saga" solver= "saga" | penalty= "l2" penalty= "l1" | class_weight= "balanced" class_weight= "none" |
| | max_iter= "150" max_iter= "180" | solver= "saga" solver= "newton-cg" | penalty= "l2" penalty= "l2" | class_weight= "balanced" class_weight= "balanced" |
| **XGBOOST** | **n_estimators= "200" (Best value)** n_estimators= "150" n_estimators= "100" n_estimators= "200" | learning_rate= "0.05" learning_rate= "0.2" learning_rate= "0.1" learning_rate= "0.01" | max_depth= "4" max_depth= "4" max_depth= "4" max_depth= "5" | random_state= "2" random_state= "2" |
| **AdaBoost** | **n_estimators= "150" (Best value)** n_estimators= "100" n_estimators= "200" n_estimators= "120" n_estimators= "250" | algorithm= "SAMME.R" algorithm= "SAMME" algorithm= "SAMME.R" algorithm= "SAMME" algorithm= "SAMME.R" | learning_rate= "0.05" learning_rate= "0.1" learning_rate= "0.05" learning_rate= "0.2" learning_rate= "0.075" | max_depth= "2" max_depth= "1" max_depth= "2" max_depth= "3" max_depth= "2" |
| **SGD** | **alpha= "0.0001" (Best value)** alpha= "0.0005" alpha= "0.001" | max_iter= "150" max_iter= "100" max_iter= "200" | loss= "hinge" loss= "log" loss= "hinge" | penalty= "l1" penalty= "l2" penalty= "l1" |
| **Ridge** | **solver="sag" (Best value)** solver="lsqr" solver="sag" | max_iter= "300" max_iter= "150" max_iter= "200" | tolerance(tol)="1e-4" tolerance(tol)="1e-4" tolerance(tol)="1e-3" | alpha= "0.015" alpha= "0.005" alpha= "0.001" |
| **RC** | **metric= "manhattan" (Best value)** metric= "euclidean'" | shrink_threshold= "0.4" shrink_threshold= "0.6" | alpha= "0.01" alpha= "0.005" | |
| **PA** | **max_iter= "200" (Best value)** max_iter= "250" max_iter= "150" | n_iter_no_change= "30" n_iter_no_change= "40" n_iter_no_change= "20" | loss= "hinge" loss= "squared_hinge" loss= "hinge" | tol=0.01 tol=0.005 tol=0.001 |
| **BPN** | **max_iter= "250" (Best value)** max_iter= "150 max_iter= "300" max_iter= "200" | hidden_layer_sizes= "100" hidden_layer_sizes= "50" hidden_layer_sizes= "100" hidden_layer_sizes= "100" | Activation function= "relu" Activation function= "relu" Activation function= "tanh" Activation function= "relu" | learning_rate_init= "0.01" learning_rate_init= "0.02" learning_rate_init= "0.005" learning_rate_init= "0.001" |

## 5.7. Results Before Hyper-Parameter Optimization

This section presents evaluation results of various machine learning methods for intrusion detection using the 1999 KDD Cup dataset, implemented with Scikit-learn. A ten-fold cross-validation was employed, dividing the dataset into ten equal parts, with nine folds for training and one for testing in each iteration. Performance metrics were averaged across folds to assess consistency. All classifiers were first evaluated using Scikit-learn's default hyper-parameters to establish baseline performance prior to hyper-parameter tuning.

Table 5. Performance comparisons of ML classifiers on the 1999 KDD Cup intrusion dataset for Network IDS before Hyper-Parameter Optimization

| Classification Algorithms | Accuracy | Precision | Recall | F1-Score | FAR | Detection Rate |
|---|---|---|---|---|---|---|
| KNN | 0.9724±0.0551 | 0.9806±0.0311 | 0.9724±0.0551 | 0.9741±0.0499 | 0.0173 | 0.9724±0.0551 |
| DT | 0.9713±0.0556 | 0.9809±0.3011 | 0.9713±0.0556 | 0.9731±0.0502 | 0.0175 | 0.9713±0.0556 |
| MNB | 0.9329±0.3869 | 0.9399±0.2870 | 0.9329±0.3868 | 0.9298±0.0389 | 0.0383 | 0.9329±0.3868 |
| BNB | 0.9473±0.0555 | 0.9594±0.0272 | 0.9473±0.0555 | 0.9487±0.0497 | 0.0338 | 0.9473±0.0555 |
| RF | 0.9714±0.0590 | 0.9811±0.0309 | 0.9714±0.0590 | 0.9733±0.0531 | 0.0188 | 0.9714±0.0590 |
| SVM | **0.9808±0.0061** | **0.9815±0.0053** | **0.9808±0.0061** | **0.9808±0.0059** | **0.0123** | **0.9808±0.0061** |
| PPN | 0.9101±0.1303 | 0.9539±0.0430 | 0.9101±0.1303 | 0.9157±0.1186 | 0.0492 | 0.9101±0.1303 |
| LR | 0.9667±0.0595 | 0.9777±0.0289 | 0.9667±0.0595 | 0.9689±0.0534 | 0.0247 | 0.9667±0.0595 |
| XGBoost | 0.9464±0.0827 | 0.9646±0.0325 | 0.9464±0.0827 | 0.9496±0.0736 | 0.0267 | 0.9464±0.0827 |
| AdaBoost | 0.9431±0.0598 | 0.9556±0.0288 | 0.9431±0.0598 | 0.9449±0.0530 | 0.0271 | 0.9431±0.0598 |
| SGD | 0.9046±0.1255 | 0.9457±0.0389 | 0.9046±0.1255 | 0.9097±0.1136 | 0.0424 | 0.9046±0.1255 |
| Ridge | 0.9495±0.0395 | 0.9562±0.0241 | 0.9495±0.0395 | 0.9499±0.0359 | 0.0339 | 0.9495±0.0395 |
| RC | 0.9487±0.0388 | 0.9549±0.0251 | 0.9487±0.0388 | 0.9488±0.0350 | 0.0357 | 0.9487±0.0388 |
| PA | 0.9443±0.0650 | 0.9576±0.0278 | 0.9443±0.0650 | 0.9462±0.0532 | 0.0373 | 0.9443±0.0650 |
| BPN | 0.9704±0.5934 | 0.9810±0.0290 | 0.9704±0.0594 | 0.9725±0.0534 | 0.0167 | 0.9704±0.0594 |

The average values and standard deviations of the classification outcomes are shown in Table 5

Table 5 summarizes the detection performance of various ML classifiers on the 1999 KDD Cup dataset using default parameters, considering accuracy and false alarm rate (FAR). SVM achieved the highest accuracy of 98.08% with the lowest FAR of 0.0123, demonstrating superior intrusion detection. KNN, RF, BPN, and DT also performed well, with accuracies above 97% and low FARs, while SGD had the lowest accuracy (90.46%) and higher FAR (0.0424). MNB recorded 93.29% accuracy with FAR 0.0383, and XGBoost, AdaBoost, and Ridge ranged between 94–95% accuracy. Fig. 19 illustrates these results, highlighting SVM's superior performance and SGD's relative ineffectiveness in its default configuration.

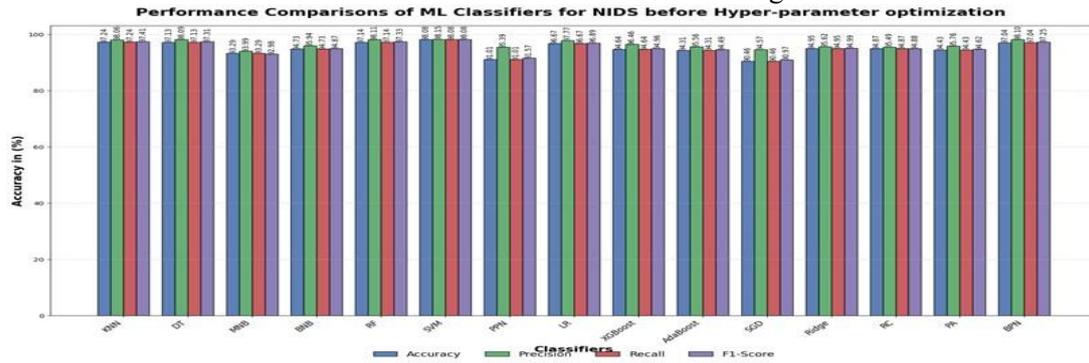

Fig. 19: Performance Comparisons of ML Classification Algorithms before hyper-parameter optimization in NIDS

## 5.8. Results After Hyper-Parameter Optimization

This section evaluates the performance of multiple ML classifiers on the 1999 KDD Cup intrusion dataset, a widely used benchmark in NIDS research. Classifiers were implemented using Scikit-learn and assessed via ten-fold cross-validation, splitting the dataset into ten folds with nine for training and one for testing per iteration to reduce variance and overfitting. To optimize the feature space and improve classifier efficiency, Recursive Feature Elimination (RFE) was applied to retain the most significant features while discarding less impactful ones. Combining RFE with cross-validation provides insights into classifier generalization and real-world applicability.

The key hyper-parameters for each classifier, including learning rate, maximum tree depth, and regularization factors, were initially set to Scikit-learn's default values. Parameters such as alpha for Ridge and C for SVM controlled model complexity and mitigated overfitting. For tree-based models like Random Forest and XGBoost, n_estimators and max_depth was adjusted to balance performance and overfitting. Hyper-parameters were further fine-tuned empirically to optimize generalization and enhance detection of multiple attack types in the 1999 KDD Cup dataset. Ten-fold cross-validation provides a reliable and unbiased evaluation of classifiers such as SVM, XGBoost, AdaBoost, RF, BNB, MNB, LR, KNN, DT, and BPN. This method enables assessment of key performance metrics, including Accuracy, Precision, Recall, F1-score, FAR, and DR, offering a comprehensive view of each classifier's effectiveness. Comparing these metrics helps identify the most suitable ML approaches for detecting and classifying network intrusions.

Table 6. Performance comparisons of ML classifiers on the 1999 KDD Cup intrusion dataset for Network IDS with Hyper Parameter Optimization

| Classification Algorithms | Accuracy | Precision | Recall | F1-Score | FAR | Detection Rate |
|---|---|---|---|---|---|---|
| KNN (K=10) (Best Value) | 0.9829±0.0649 | 0.9899±0.0409 | 0.9829±0.0649 | 0.9847±0.0596 | 0.0148 | 0.9829±0.0649 |
| DT | 0.9812±0.05 | | | | 0.0162 | 0.9812±0.0589 |

| | | | | | |
|---|---|---|---|---|---|
| **(Best Value)** | 89 | 0.9904±0.3088 | 0.9812±0.0589 | 0.9821±0.0556 | | |
| **MNB**<br>**(Best Value)** | 0.9568±0.0869 | 0.9638±0.2980 | 0.9568±0.0869 | 0.9537±0.0491 | 0.0345 | 0.9568±0.0869 |
| **BNB**<br>**(Best Value)** | 0.9612±0.0682 | 0.9732±0.0372 | 0.9612±0.0682 | 0.9626±0.0572 | 0.0301 | 0.9612±0.0682 |
| **RF**<br>**(Best Value)** | 0.9827±0.0627 | 0.9899±0.0498 | 0.9827±0.0627 | 0.9849±0.5890 | 0.0169 | 0.9827±0.0627 |
| **SVM**<br>**(Best Value)** | **0.9912±0.0097** | **0.9919±0.0088** | **0.9912±0.0097** | **0.9912±0.0095** | **0.0091** | **0.9912±0.0097** |
| **PPN**<br>**(Best Value)** | 0.9477±0.0410 | 0.9563±0.0573 | 0.9477±0.0402 | 0.9533±0.1254 | 0.0467 | 0.9477±0.0402 |
| **LR**<br>**(Best Value)** | 0.9769±0.0695 | 0.9849±0.0398 | 0.9769±0.0695 | 0.9789±0.0634 | 0.0220 | 0.9769±0.0695 |
| **XGBoost**<br>**(Best Value)** | 0.9787±0.0476 | 0.9850±0.0308 | 0.9787±0.0476 | 0.9811±0.0420 | 0.0237 | 0.9787±0.0476 |
| **AdaBoost**<br>**(Best Value)** | 0.9776±0.0594 | 0.9879±0.0292 | 0.9776±0.0594 | 0.9897±0.0534 | 0.0241 | 0.9776±0.0594 |
| **SGD**<br>**(Best Value)** | 0.9488±0.0498 | 0.9565±0.0368 | 0.9488±0.0498 | 0.9511±0.1386 | 0.0396 | 0.9488±0.0498 |
| **Ridge**<br>**(Best Value)** | 0.9618±0.0508 | 0.9689±0.0346 | 0.9618±0.0508 | 0.9625±0.0389 | 0.0310 | 0.9618±0.0508 |
| **RC**<br>**(Best Value)** | 0.9561±0.0462 | 0.9626±0.0296 | 0.9561±0.0462 | 0.9562±0.0410 | 0.0330 | 0.9561±0.0462 |
| **PA**<br>**(Best Value)** | 0.9533±0.0699 | 0.9616±0.0375 | 0.9533±0.0699 | 0.9554±0.0632 | 0.0352 | 0.9533±0.0699 |

| **BPN (Best Value)** | 0.9821±0.0938 | 0.9915±0.0392 | 0.9821±0.0938 | 0.9830±0.0638 | 0.0152 | 0.9821±0.0938 |
|---|---|---|---|---|---|---|

The average values and standard deviations of the classification outcomes are shown in Table 6

Table 6 highlights SVM as the top-performing classifier, achieving 99.12% Accuracy, 99.19% Precision, 99.12% Recall, 99.12% F1-score, and the lowest FAR of 0.0091. KNN and BPN closely followed with accuracies of 98.29% and 98.21% and low FARs of 0.0148 and 0.0152. DT and RF showed similar reliability, with accuracies of 98.12% and 98.27% and FARs of 0.0162 and 0.0169. Ensemble methods, XGBoost (97.87%) and AdaBoost (97.76%), delivered strong performance, though below SVM. Lightweight classifiers, BNB (96.12%) and MNB (95.68%), were suitable for resource-limited scenarios but had higher FARs (0.0301 and 0.0345). Logistic Regression achieved 97.69% accuracy with FAR 0.0220. PA, SGD, and Perceptron underperformed, with detection rates below 95.50%, limiting their suitability for critical intrusion detection tasks. Figs. 20–34 present line graphs illustrating classifier performance after hyper-parameter tuning across various metrics.

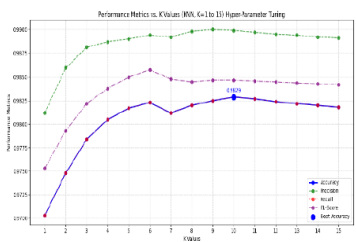

Fig 20. Hyper-Parameter Tuning for KNN

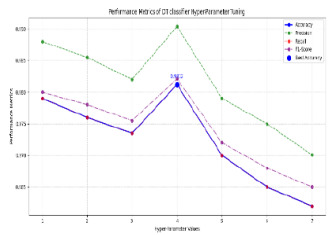

Fig 21. Performance of DT

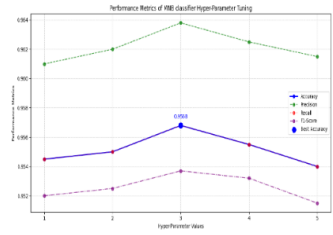

Fig 22. Performance of MNB

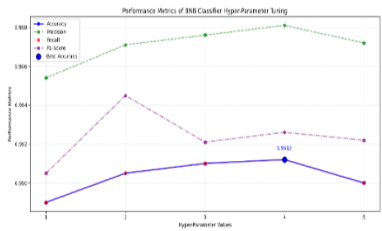

Fig 23. Performance of BNB

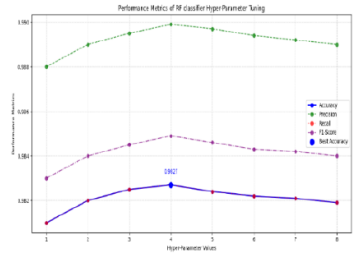

Fig 24. Performance of RF

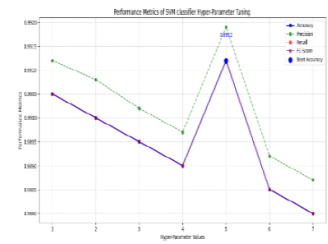

Fig 25. Performance of SVM

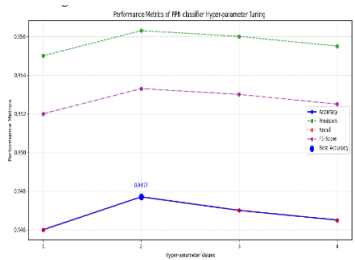

Fig 26 Performance of PPN

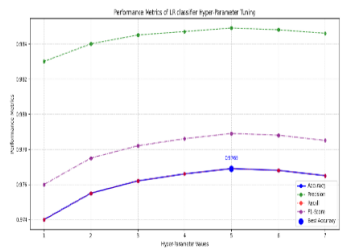

Fig 27. Performance of LR

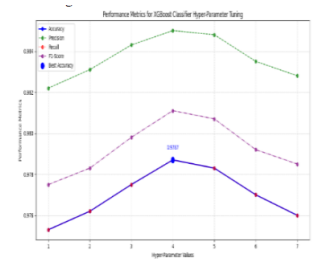

Fig 28. Performance of XGBoost

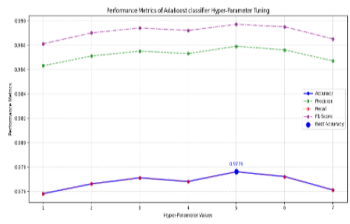
Fig 29. Performance of AdaBoost

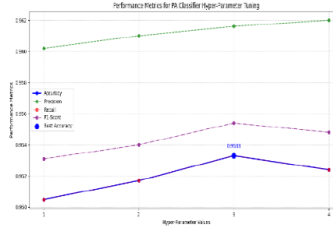
Fig 30. Performance of PA

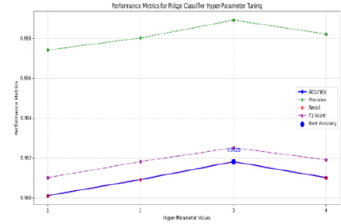
Fig 31. Performance of Ridge

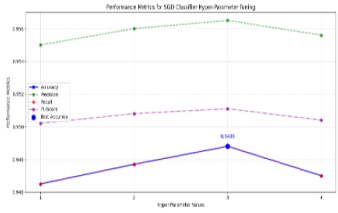
Fig 32. Performance of SGD

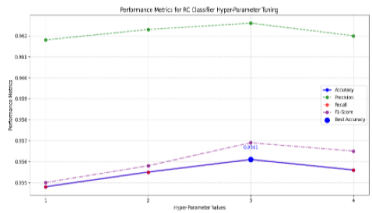
Fig 33. Performance of RC

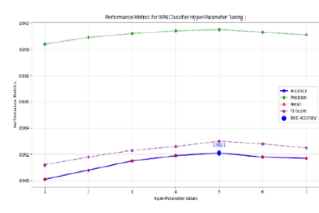
Fig 34. Performance of BPN

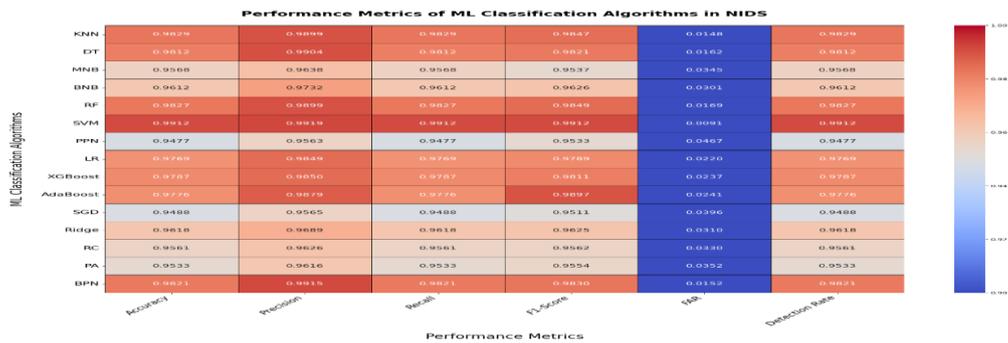

Fig 35. Heat map representation of prediction accuracy for different machine learning algorithms in NIDS

Fig. 35 presents a heatmap comparing 15 ML classifiers across Accuracy, Precision, Recall, F1-score, FAR, and Detection Rate. The color gradient highlights detection performance and false alarm control, helping identify the most effective classifier for network intrusion detection.

## 5.9. Analysis of ROC Curves for Machine Learning Classifiers on the 1999 KDD Cup Intrusion Data

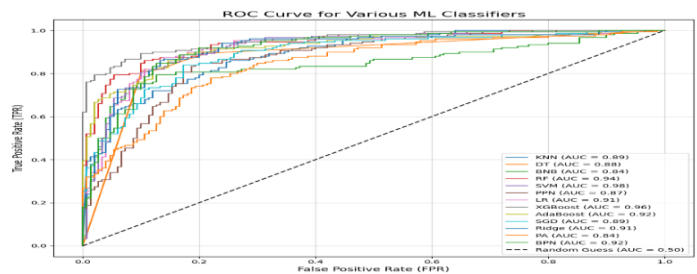

Fig 36. Performance Evaluation Using ROC Curves: ML Classifiers on the 1999 KDD Cup Intrusion Dataset

Fig. 36 shows ROC analysis on the KDD Cup 1999 dataset, where SVM, BPN, and RF achieved the highest AUC (~0.98). KNN, DT, and BNB followed (0.89–0.94), while XGBoost and AdaBoost performed moderately (0.86–0.88). All models surpassed the random baseline, confirming effective intrusion detection.

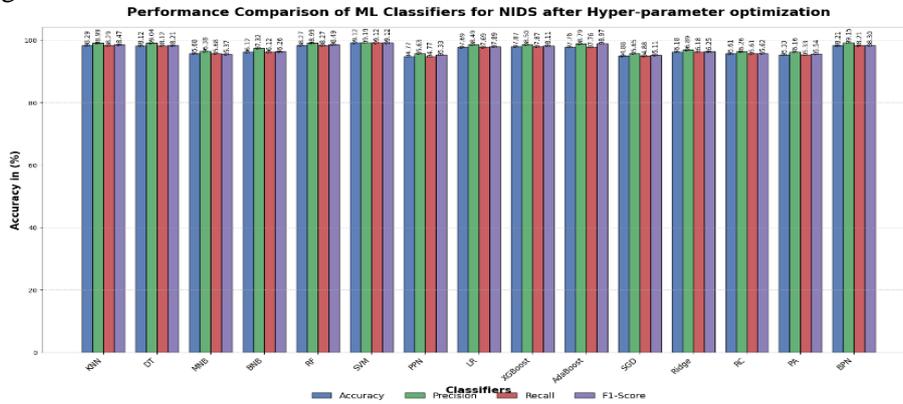

Fig 37. Performance Comparisons of ML Classification Algorithms with hyper-parameter optimization in NIDS

Fig. 37 presents a bar chart comparing classifiers across Accuracy, Precision, Recall, F1-score, and FAR. SVM achieves the highest performance, with 99.12% accuracy, strong Precision, Recall, and F1-score, and a minimal FAR, highlighting its effectiveness in detecting network intrusions on the 1999 KDD Cup dataset. KNN, RF, BPN, DT, and LR also show strong results, while ensemble methods like XGBoost and AdaBoost perform well but do not surpass SVM. Linear models such as Ridge and SGD exhibit moderate performance, reflecting challenges in handling dataset complexity. These results underscore SVM's superiority for NIDS and provide guidance for selecting and tuning classifiers for robust intrusion detection.

## 6. CONCLUSION AND FUTURE SCOPE

Detecting network intrusions is critical for maintaining cybersecurity, and machine learning (ML) has proven effective in identifying malicious activities within network traffic. Supervised ML algorithms enable systems to distinguish legitimate from suspicious behavior, enhancing protection against evolving threats. Using the 1999 KDD Cup intrusion dataset, this study applied hyper-parameter tuning to optimize classifiers. SVM, KNN, RF, and XGBoost achieved high accuracy and reliable detection rates, while Perceptron (PPN) and Stochastic Gradient Descent (SGD) performed less effectively. Classifiers such as Naïve Bayes, Ridge, and Passive Aggressive showed moderate performance, highlighting variability in algorithm effectiveness for NIDS. As cyber threats evolve, future research will focus on advanced techniques capable of handling large-scale, dynamic datasets. Unsupervised methods, including K-means, OC-SVM, Isolation Forest, DBSCAN, and Autoencoders, are essential for detecting novel attacks, such as zero-day threats, without labeled data. Hybrid approaches combining multiple learning paradigms can improve adaptability and detection accuracy. Moreover, integrating Explainable AI (XAI) will enhance transparency and interpretability, fostering trust in real-world deployment. These advancements promise more adaptive, scalable, and robust intrusion detection systems capable of mitigating increasingly sophisticated cyber threats.

**AUTHORS**

**SUDHANSHU SEKHAR TRIPATHY** (Member, IEEE) received his Master of Computer Applications (MCA) degree in 2017 from Gandhi Institute for Technology (GIFT), affiliated with Biju Patnaik University of Technology (BPUT), Odisha, India. He is currently pursuing a Ph.D. in Computer Science and Engineering at C.V. Raman Global University, Bhubaneswar, Odisha. He has three years of industry experience and five years of academic teaching experience. His research work has been published in two Scopus-indexed and one Web of Science-indexed journal and two UGC CARE Group 1 journals, focusing on cybersecurity with an emphasis on Network Intrusion Detection Systems. His research interests include machine learning, deep learning, network security, zero-day attack detection, and intrusion detection systems.

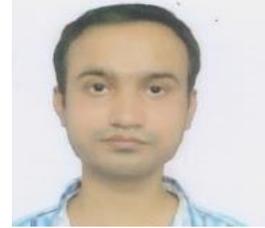

**BICHITRANANDA BEHERA** (Member, IEEE) received his Ph.D. in 2022 from Pondicherry University, Puducherry, India. He is currently working as an assistant professor at C.V. Raman Global University, Bhubaneswar, Odisha, India. He has published numerous research articles in reputed journals and international conferences. His current research interests include machine learning, data mining, natural language processing (NLP), cybersecurity, and soft computing. He is actively engaged in the academic community and serves as a reviewer for various reputable journals and publications.

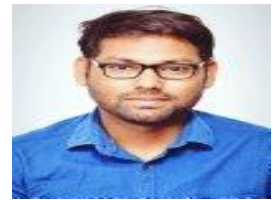